\documentclass[amsmath, amssymb, superscriptaddress, twocolumn, prb]{revtex4-2}

\usepackage{graphicx}
\usepackage{dcolumn}
\usepackage{bm}
\usepackage{bbm}
\usepackage{ulem}
\usepackage{cases}
\usepackage{xcolor} 
\usepackage{hyperref}
\usepackage{epstopdf}
\usepackage[mathscr]{eucal}

\begin{document}

\title{Electric circuit analog of Landau-Zener tunneling using time-varying elements} 

\author{Enhong Cheng}
\thanks{They contribute equally to this work.}
\affiliation{School of Physics, South China Normal University, Guangzhou 510006, China}

\author{Zheng Lian}
\thanks{They contribute equally to this work.}
\affiliation{School of Physics, South China Normal University, Guangzhou 510006, China}

\author{Zezhou Chen}
\affiliation{School of Physics, South China Normal University, Guangzhou 510006, China}

\author{Li-Jun Lang}
\email{ljlang@scnu.edu.cn}
\affiliation{School of Physics, South China Normal University, Guangzhou 510006, China}
\affiliation{Guangdong Provincial Key Laboratory of Quantum Engineering and Quantum Materials, South China Normal University, Guangzhou 510006, China}
\affiliation {Guangdong-Hong Kong Joint Laboratory of Quantum Matter, South China Normal University, Guangzhou 510006, China}

\date{\today}    

\begin{abstract}

Landau-Zener tunneling (LZT) is a fundamental dynamical phenomenon, ubiquitous in various quantum systems. 
Here, we propose a time-varying electric circuit to address the question of whether the quantum LZT can occur in classical systems. 
Although the underlying differential equation of motion is quite different from the Schr\"odinger equation and the instantaneous frequency spectrum of the proposed circuit is not linear, the probability of the LZT in circuits (circuit LZT for short), based on our generalized definition for norm-unconserved systems, still follows the laws of the LZT in quantum systems, codetermined by the linear sweeping rate $\alpha'$ and the frequency gap $\Delta$, i.e., approaching the analytical value $\exp(-\pi\Delta^2/2\alpha')$, regardless of whether the coupling is reciprocal or nonreciprocal.
The deep relationship between the circuit LZT and its quantum counterpart can be established through a linearization and block-diagonalization process.
Our proposal provides a general method for simulating time-dependent quantum models using time-varying electric circuits, which has been lacking in previous studies, and paves the way for studying more complicated LZT and other dynamical phenomena in circuits and other classical systems.

\end{abstract}

\maketitle

\section{Introduction \label{SEC.I}}

Landau-Zener tunneling (LZT) is a fundamental dynamical phenomenon in quantum systems, describing quantum tunneling between two avoided crossing levels at a certain sweeping rate \cite{Landau_1932_PZS,Zener_1932_PRSA,Stuckelberg_1932_HPA,MajoranaZ_1932_IINC}. 
It has been widely demonstrated in various experimental platforms, such as superconducting qubits \cite{Sillanpaa_Pertti_2006_PRL,Berns_Orlando_2008_Nature,Sun_Han_2010_NC,Tan_Zhu_2014_PRL,Sun_Han_2015_SciRep}
, nitrogen-vacancy centers \cite{Fuchs_Awschalom_2009_Science}, quantum dots \cite{Petta_Gossard_2010_Science}, Bose-Einstein condensates \cite{Morsch_Arimondo_2001_PRL,Cristiani_Ennio_2002_PRA,Zenesini_Wimberger_2008_NJP,Zenesini_Arimondo_2009_PRL}, and Rydberg atoms \cite{Zhang_Liu_2018_PRL}. 
It is also the basis for studying LZTs under more complicated settings \cite{Ivakhnenko_Nori_2023_PR}, such as multiple level systems \cite{Sinitsyn_2002_PRB,Shytov_2004_PRA,Volkov_Ostrovsky_2004_JPB,Sinitsyn_2013_PRA,Sinitsyn_2014_PRA,Li_Sinitsyn_2017_PRA,Malla_Raikh_2017_PRB}, nonlinear or interacting systems \cite{Wu_Niu_2000_PRA,Zobay_Barry_2000_PRA,Liu_Niu_2002_PRA,Witthaut_Korsch_2006_PRA,Wu_Liu_2006_PRL,Wang_Liu_2006_PRA,Smith_Cohen_2009_PRL,Chen_Altman_2011_NP,Kasztelan_Orso_2011_PRL,Zhang_Chen_2019_PRA,Guan_Blume_2020_PRL}, open systems \cite{Gefen_Caldeira_1987_PRB,Wubs_Kayanuma_2006_PRL,Ashhab_2016_PRA}, and systems with nonlinear sweeps \cite{Garanin_Schilling_2002_PRB,Dou_Fu_2018_PRA}.
Recently, the study of LZT has also been extended to non-Hermitian systems \cite{Akulin_Schleich_1992_PRA,Avishai_Band_2014_PRA,Torosov_Vitanov_2017_PRA,Shen_Wu_2019_PRA,He_Jones_2021_PRA,Wang_Liu_2022_PRA,Wang_Fu_2023_NJP,Ivakhnenko_Nori_2023_PR,Kivela_Paraoanu_2024_PRR}.

Although LZT is a prototypical quantum phenomenon, one may be curious whether it can occur in classical systems since the underlying mechanisms are quite different, as manifested by the differential equations of motion that differ from the Schr\"odinger equation of quantum systems.
In addition, using classical systems such as photonic crystals \cite{Raghu_Haldane_2008_PRA, Wang_Soljacic_2009_Nature, Rechtsman_Szameit_2013_Nature, Hafezi_Taylor_2013_NatPhotonics, Yan_Lu_2018_NP, Ozawa_Carusotto_2019_RMP}, phononic crystals \cite{Hussein_Ruzzene_2014_AMR, Wang_Bertoldi_2015_PRL, Xiao_Chan_2017_PRL, Li_Liu_2018_NP, Li_Li_2020_PRR, Li_Li_2021_PRR}, and mechanical oscillators \cite{Kane_Lubensky_2014_NP, Chen_Vitelli_2014_PNAS, Nash_Irvine_2015_PNAS,Chaunsali_Yang_2017_PRL, Ivakhnenko_Nori_2018_SR,Brandenbourger_Coulais_2019_NC, Ghatak_Coulais_2020_PNAS}
to simulate quantum phenomena has emerged as a hot topic due to their economic costs, mature techniques, and flexible manipulations. 
Among them, the electric circuit has become a particularly powerful competitor for simulating quantum tight-binding models because of its limitless capability to construct networks of any dimension and any boundaries \cite{Zhao_2018_AoP,Luo_Yu_2019_APS,Liu_Cui_2021_CO,Xu_Lang_2023_APS,Sahin_Lee_2025_APL}.
Although there have been considerable electric-circuit proposals \cite{Albert_Jiang_2015_PRL,Ezawa_2018_PRB,Ezawa_2019_PRB,Hofmann_Thomale_2019_PRL,Zhang_Franz_2020_PRL,Lang_Liang_2021_PRB,Rafi-Ul-Islam_Mansoor_2021_NJP,Dong_Roy_2021_PRR} and experiments \cite{Schindler_Kottos_2012_JoPA, Ningyuan_Simon_2015_PRX, Lee_Thomale_2018_CP, Imhof_Thomale_2018_NP, Helbig_Kiessling_2019_PRB, Liu_Zhang_2019_Research, Bao_Zhang_2019_PRB,Lee_Thomale_2020_NC, Wu_Liu_2020_PRB, Helbig_Thomale_2020_NP,Zou_Zhang_2021_NC,Kotwal_Dunkel_2021_PNAS,Yang_Chong_2021_SCPM,Zhang_Zhang_2021_PRL,Wu_Yu_2022_NE,Wang_Yu_2023_PRL,Stegmaier_Upreti_2024_PRR,Stegmaier_Thomale_2024_arxiv,Zhang_Zhang_2025_NC,Zhang_Jia_2025_NC} to simulate quantum physics, most of them focus on time-independent models from the perspective of the spectrum based on the circuit Hamiltonian or Laplacian \cite{Wu_2004_JPAMG,Tzeng_Wu_2006_JPAMG,Lee_Thomale_2018_CP}, with limited attention given to time-dependent ones \cite{Stegmaier_Upreti_2024_PRR,Stegmaier_Thomale_2024_arxiv,Zhang_Zhang_2025_NC,Zhang_Jia_2025_NC}.
This is mainly due to the fundamental differences: Kirchhoff's equation of circuits usually involves a second-order derivative with respect to time, rather than the first-order one of the time-dependent Schr\"odinger equation; additionally, the dynamical quantities in circuits are always real, not complex like quantum wavefunctions.

Therefore, to investigate the analog of LZT in classical systems, especially in electric circuits and, meanwhile, to find a general method for simulating time-dependent quantum models, we propose a time-varying electric circuit to demonstrate the circuit analog of LZT (circuit LZT for short).
We find that the circuit LZT not only qualitatively follows the rules of LZT in quantum systems (quantum LZT for short) but also quantitatively provides the same probability as its quantum counterpart, based on our generalized definition of probability for norm-unconserved systems. 
Although our circuit model has a nonlinear instantaneous frequency spectrum, the probability can be well explained by our linearized and block-diagonalized model, which establishes a mapping between the circuit and quantum LZTs, approaching the analytical value $\exp(-\pi\Delta^2/2\alpha')$ if the linearized sweeping rate $\alpha'$ and the frequency gap $\Delta$ are identical in the two mapped systems. 
Based on this mapping, our proposal provides a general method for simulating quantum dynamical phenomena using time-varying electric circuits.

The remaining parts are organized as follows. 
In Sec. \ref{Sec.II}, we propose a time-varying electric circuit using linearly varying capacitors for circuit LZT. 
In Sec. \ref{Sec.III}, the numerical results of circuit LZT for the reciprocal case are shown.  
In Sec. \ref{sec:relation}, we provide an explanation by mapping the circuit LZT to its quantum counterpart through a linearization and block-diagonalization process. 
In Sec. \ref{sec_v}, we generalize the circuit LZT to the nonreciprocal case.
Conclusions are given in Sec. \ref{Sec:conclusion}.

\section{Setup of the electric circuit} \label{Sec.II}

\begin{figure}[tb]	
    \includegraphics[width=1\linewidth]{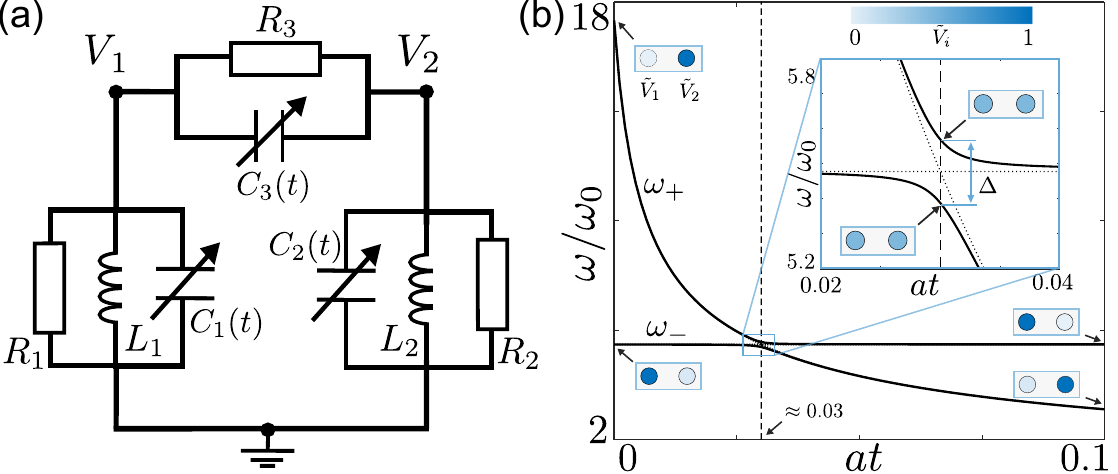}
    \caption{(a) The circuit diagram for circuit LZT, composed of resistors (R), inductors (L), and time-varying capacitors (C). $V_{1,2}(t)$ are the voltages of node 1 and 2.
	(b) The instantaneous eigenfrequencies $\omega_\pm(t)$ (solid lines) and the channel frequencies $\omega_{1,2}(t)$ (dotted lines) with parameters $(r,b)=(30,1/300)$, independent of $v$. 
    The dashed vertical line indicates the crossing point $at_c\approx0.03$, at which the frequency gap $\Delta\approx 0.18$.
    The rectangles including two circles demonstrate the voltage distributions $\tilde{V}_{\pm,i}(t)$ on the two nodes for the corresponding instantaneous eigenmodes.
    }\label{fig1}
\end{figure}

The physical nature of LZT in quantum systems can be exactly captured by the time-dependent Schr\"odinger equation of a two-level system with a linear sweep \cite{Zener_1932_PRSA}
\begin{eqnarray}\label{schEQ}
	i\frac{d}{d t}|\psi(t)  \rangle= H(t)|\psi(t)  \rangle,
\end{eqnarray}
where 
\begin{eqnarray}\label{LZ_Ham}
	H(t)  = \frac{1}{2}
	\begin{bmatrix}	
    ~\alpha t~&~\nu \\
    ~\nu &~~~	-\alpha t
    \end{bmatrix}
\end{eqnarray}
is the Hamiltonian matrix, and 
$|\psi(t) \rangle= [\psi_1(t), \psi_2(t)]^T$ is the state vector at time $t$, with $\psi_{1,2}(t)$ being the amplitudes of the state in channels 1 and 2, respectively;
$\alpha t$ describes the linear sweep of channel potentials with the sweeping rate $\alpha$, and $\nu$ is the strength of inter-channel coupling.

To explore LZT in electric circuits, we set up a time-varying circuit that is composed of two RLC loops as the two channels and one RC loop as the interchannel coupling, as shown in Fig. \ref{fig1}(a), where $L_i$, $R_i$, and $C_i(t)$ label the corresponding inductance, resistance, and time-varying capacitance, respectively.
The usage of time-varying capacitors is to realize the sweep of channel frequencies for circuit LZT; one may also use time-varying inductors or resistors instead. 
Considering the note voltages $V_{1,2}(t)$ in Fig. \ref{fig1}(a) as the dynamical variables similar to $\psi_{1,2}(t)$ in Eq. \eqref{schEQ}, according to Kirchhoff’s laws, the corresponding differential equation can be cast into a compact form
	\begin{equation}\label{EQvoltage}
		\mathcal{C}_C(t)\ddot{\mathcal{V}}(t)+\mathcal{C}_R(t)\dot{\mathcal{V}}(t)+\mathcal{C}_L(t)\mathcal{V}(t)=0, 
	\end{equation}
where the vector $\mathcal{V}(t) = [V_1(t),V_2(t)]^T$ and the time-dependent coefficient matrices
\begin{equation}\label{coefficientMatrix}
\begin{aligned}
	\mathcal{C}_C(t) &= 
        \begin{bmatrix}
        -C_3-C_1&C_3 \\
        C_3&-C_3-C_2
        \end{bmatrix},\\
	\mathcal{C}_R(t) &=
        \begin{bmatrix}
        -C_R^{(1)}& 2\dot{C}_3+R_3^{-1}\\
        2\dot{C}_3+R_3^{-1}&-C_R^{(2)}
        \end{bmatrix},\\
	\mathcal{C}_L(t) &=
        \begin{bmatrix}
        -C_L^{(1)} & \ddot{C}_3\\
        \ddot{C}_3 &-C_L^{(2)}
        \end{bmatrix},
\end{aligned}
\end{equation}
with $C_R^{(i)} = 2(\dot{C}_i+\dot{C}_3)+R_i^{-1}+R_3^{-1}$ and $C_L^{(i)} = \ddot{C}_i+\ddot{C}_3+L_i^{-1}$.
The dots above the quantities represent the time derivatives.
The detailed derivation can be found in Appendix \ref{Anote2}.

To get close to the form of Eq. \eqref{schEQ}, 
we choose the following parametrization:
\begin{equation}
\begin{aligned}
	\frac{C_1}{C_0} &= \frac{1}{rv}+\frac{1-vr}{vr^2}\left(a t+b\right), & R_1C_0& = \frac{vr^2}{2a (vr-1)},\\
	\frac{C_2}{C_0} & = \frac{vr-1}{r}\left(a t+b\right), & R_2C_0 & = \frac{r}{2a (1-vr)},\\
	\frac{C_3}{C_0} & = \frac{a t+b}{r}, & R_3C_0 &= -\frac{r}{2a},\\
	\frac{L_1}{L_0} &= v,  &\frac{L_2}{L_0} &= \frac{1}{v},
\end{aligned}
\label{param}
\end{equation}
where $v$, $r$, and $b$ are positive dimensionless parameters, and $C_0$ and $L_0$ are the reference capacitance and inductance, respectively; the nonnegative parameter $a$, with an inverse-of-time dimension, controls the change rate of capacitors (see Appendix \ref{AnoteTVC}).
In Eq. \eqref{param}, the capacitance or resistance may be negative, and one may resort to the schemes of realizing negative elements in circuits \cite{Schindler_Kottos_2012_JoPA}.
After parametrization in Eq. \eqref{param}, Eq. \eqref{EQvoltage} can be reduced to
\begin{equation}\label{ddv}
	\ddot{\mathcal{V}}(t) = -\mathcal{H}(t)\mathcal{V}(t),
\end{equation}
where the coefficient matrix (dubbed the circuit Hamiltonian)
\begin{equation}\label{CcCl}
    \mathcal{H}(t)  = 
    \mathcal{H}_0+\mathcal{H}_c 
    \equiv
    \begin{bmatrix}	
    ~\omega_1^2 &~0 \\
    ~0 & ~\omega_2^2	
    \end{bmatrix}
    +\omega_0^2
    \begin{bmatrix}	
    ~0 &~v \\
    ~v^{-1} & ~0
    \end{bmatrix}
\end{equation}
takes the role of $H(t)$ in Eq. \eqref{schEQ} to determine the dynamics of node voltages; the diagonal part
$\mathcal{H}_0$ represents the uncoupled circuit Hamiltonian, with $\omega_1\equiv\omega_0\sqrt{r}\ge 0$ and $\omega_2\equiv\omega_0\sqrt{r^{-1}+(at+b)^{-1}}\ge0$ being the frequencies of channels 1 and 2, respectively,
and the off-diagonal part $\mathcal{H}_c$ couples the two channels.
The reference frequency $\omega_0\equiv 1/\sqrt{L_0C_0}$ is set to 1 as the unit of frequency in the following.
The detailed derivation can be referred to in Appendix \ref{Anote2}.

Under these settings, the differential Eq. \eqref{ddv} can describe an analog of LZT in this classical circuit through node voltages, since two channel frequencies $\omega_{1,2}$ cross each other in a certain time $t_c$, at which a finite frequency gap opens due to the interchannel coupling $v^{\pm1}$, as shown in Fig. \ref{fig1}(b).
Therefore, the question of LZT in the circuit can be similarly raised as the quantum counterpart: How many fractions can one channel mode transfer to the other when the two channels are coupled during the process in which the two channel frequencies first converge, then cross, and finally diverge in time?  

Despite similar physical processes, there are fundamental differences between the circuit system and its quantum counterpart. Compared with Eq. \eqref{schEQ} in detail, Eq. \eqref{ddv} is a second-order rather than a first-order differential equation, where the node voltages, as dynamical variables, are required to be real. 
Moreover, although the capacitors in the circuit change linearly over time, the diagonal terms $\omega_{1,2}^2$ in $\mathcal{H}(t)$ as a function of time are not as linear as $\pm\alpha t/2$ in $H(t)$, due to the complicated time derivatives for time-varying elements in the derivation of Eq. \eqref{ddv} (see Appendix \ref{Anote2} for details).
These differences show that the circuit LZT does not seem to be a simulation of quantum LZT. Therefore, it is necessary to investigate it independently. 
By the way, in our design, the nonreciprocal coupling, represented by unequal terms $v^{\pm 1}$ in Eq. \eqref{CcCl}, is generally proposed with different inductors $L_{1,2}$, and thus, the nonreciprocal circuit LZT can also be investigated here.

Two instantaneous eigenfrequencies $\omega_\pm(t)$ of the circuit at time $t$, as shown in Fig. \ref{fig1}(b), can be calculated using the following eigenvalue equation:
\begin{equation}
	\mathcal{H}(t) \mathcal{V}_\pm(t) = \omega_\pm^2(t)\mathcal{V}_\pm(t),
\end{equation}
where $\mathcal{V}_\pm(t)=[V_{\pm,1}(t),V_{\pm,2}(t)]^T$ are the corresponding eigenvectors, representing two instantaneous eigenmodes; the subscript $\pm$ labels the upper ($+$) and lower ($-$) branches of instantaneous eigenfrequencies.
It is necessary for the eigenvalues $\omega_\pm^2(t)$ to be nonnegative for the stability of the dynamics, which leads to the condition $at+b>0$ (see the proof in Appendix \ref{Anote2}).
Therefore, we consider the dynamical process from $t=0$ to $+\infty$, with $b$ being a small positive number to prevent divergence at $t=0$ in Eq. \eqref{CcCl}. 
The crossing time of the two channel frequencies is determined by
\begin{equation}\label{gmc}
    t_c= \dfrac{r}{a(r^2-1)}-\frac{b}{a},
\end{equation}
at which the coupling opens a frequency gap
\begin{equation}\label{freq_gap}
    \Delta \equiv \omega_+(t_c)-\omega_-(t_c)
    =\sqrt{r+1}-\sqrt{r-1}.
\end{equation}
The crossing must occur during the dynamical process, i.e., $t_c>0$, which requires $r>1$. As a result, the frequency gap $\Delta$ is guaranteed to be real and positive.
Moreover, to ensure that the differences between channel frequencies are much larger than the coupling when $t\rightarrow0$ and $+\infty$, the parameters in Eq. \eqref{param} must satisfy the following condition:
\begin{equation}\label{paramconst}
	r-\frac{1}{r}\gg \max\Big\{v,\frac{1}{v}\Big\}.
\end{equation}
In our numerical calculation, we use $(r,b) = (30,1/300)$ and $v\ge 1$. Correspondingly, $at_c\approx0.03$ and $\Delta\approx0.18\omega_0$.

Since the matrix $\mathcal{H}(t)$ and its eigenvalues $\omega^2_\pm(t)$ are all real, the instantaneous eigenvectors $\mathcal{V}_\pm(t)$ must also be real. For each eigenvector, the voltage distribution on the two nodes can be reflected by normalized node voltages:
\begin{equation}\label{normalizedVoltages}
	\tilde{V}_{\pm,i}(t)=\frac{V_{\pm,i}(t)}{\sqrt{V^2_{\pm,1}(t)+V^2_{\pm,2}(t)}}.
\end{equation}
In Fig. \ref{fig1}(b), one can observe that the voltages of the instantaneous eigenmodes are primarily located at one node at $t\rightarrow 0$ and $\infty$, while near the crossing point $t_c$, the voltages are almost equally shared between the two nodes.

The instantaneous spectrum, along with voltage distributions, guides us in investigating the circuit LZT.
To this end, we initially inject a voltage into one of the two nodes at $t=0$ and then observe the voltage transfer between the two nodes at late times.
In the following sections, we first consider the reciprocal case (i.e., $v=1$) and then the nonreciprocal case (i.e., $v> 1$).

\section{Circuit LZT: reciprocal case} \label{Sec.III}

\begin{figure}[tb]
	\centering
	\includegraphics[width=1 \linewidth]{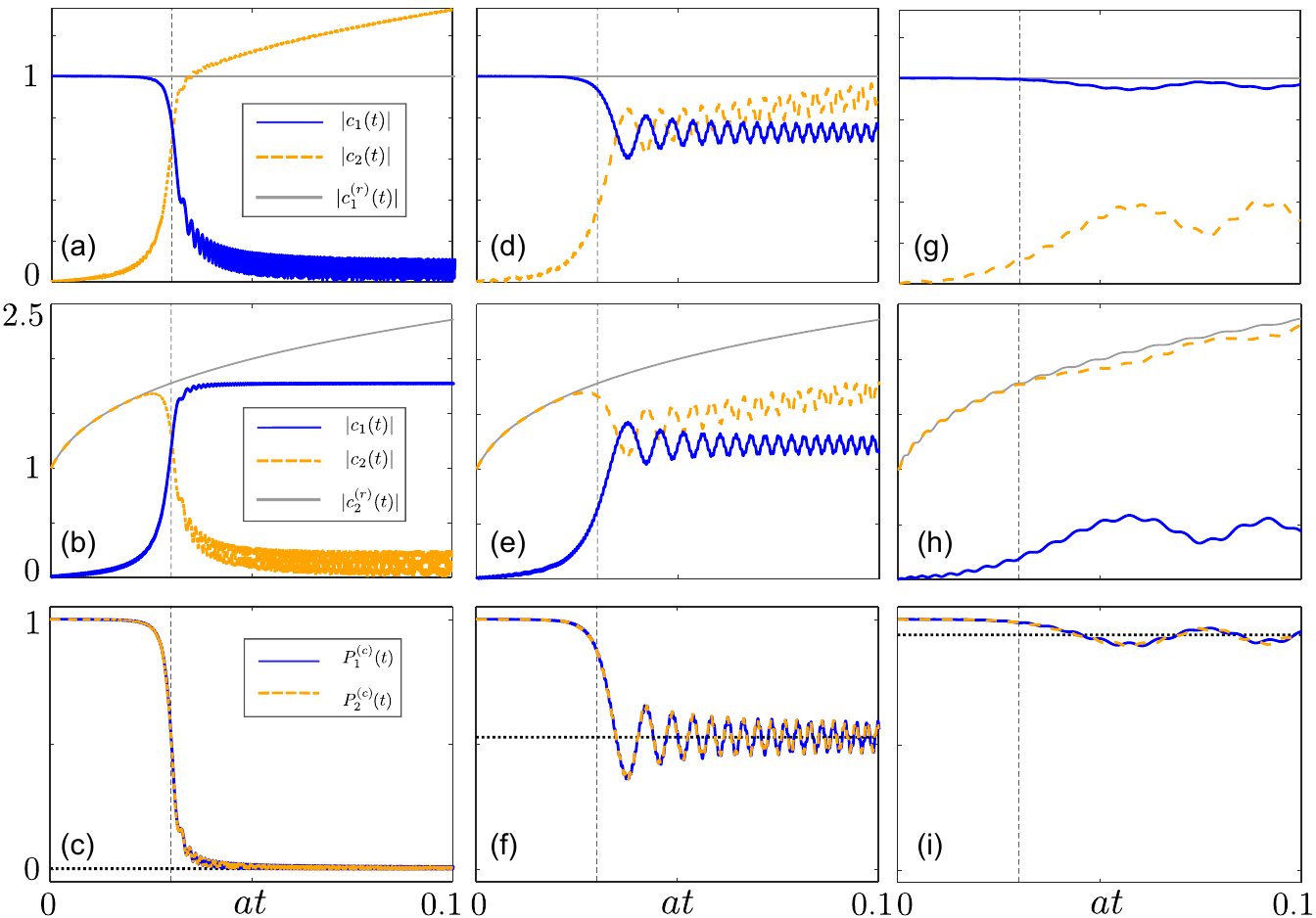}
	\caption{Reciprocal case with $v=1$. (a) and (b) Time evolution of the moduli of coefficients $|c_{1,2}(t)|$ (blue solid and orange dashed lines) for sweeping rate $a=10^{-4}$, calculated by Eqs. \eqref{eq:1stode} and \eqref{eq:proj2channel} with initial values (a) $\mathcal{U}^T(0)=[1,0,0,0]$ and (b) $\mathcal{U}^T(0)=[0,1,0,0]$.    
    The gray solid lines are the moduli of reference coefficients $|c^{(r)}_{1,2}(t)|$ (see the text for the definition).
    (c) Time evolution of remaining probabilities $P_{1,2}^{(c)}(t)$ of channels 1 and 2 corresponding to (a) and (b), respectively, calculated by Eq. \eqref{Pt}. 
    The horizontal black dotted line denotes the value of $\exp{(-\pi/2\tilde\alpha')}$ for comparison, with the dimensionless sweeping rate $\tilde\alpha'\approx0.24$.
    (d)--(f) The same as (a)--(c), except with $a=10^{-3}$ and $\tilde\alpha'\approx2.4$.
    (g)--(i) The same as (a)--(c), except with $a=10^{-2}$ and $\tilde\alpha'\approx24$.
    The parameters $r$ and $b$ are set the same as those in Fig. \ref{fig1}. 
    The vertical gray dashed lines in all figures indicate the crossing time $t_c\approx0.03/a$ in Eq. \eqref{gmc}.  
	}
	\label{fig2}
\end{figure}

One should be reminded of the mathematical differences in the differential equations for the circuit and its quantum counterpart. 
To launch the time evolution of the second-order differential Eq. \eqref{ddv}, one should first set real values of $\mathcal{V}(0)$ and $\dot{\mathcal{V}}(0)$ simultaneously, where we regard voltages as dimensionless quantities for simplicity.
In the following calculation, we simply set $\dot{\mathcal{V}}(0)=0$ unless otherwise noted. The case of $\dot{\mathcal{V}}(0)\ne0$ can be found in Appendix \ref{asec:nonzero_dotv}.

Mathematically, it is more convenient to evaluate the dynamics by rewriting the second-order differential Eq. \eqref{ddv} in the form of a first-order one
\begin{equation}\label{eq:1stode}
    \dot{\mathcal{U}}(t)\equiv\frac{d}{dt}
    \begin{bmatrix}	
    \mathcal{V} \\
    \dot{\mathcal{V}}	
    \end{bmatrix}
	= 
    \begin{bmatrix}	
    \mathbb{O} & \mathbb{I} \\
    -\mathcal{H} & \mathbb{O}	
    \end{bmatrix}
    \begin{bmatrix}	
    \mathcal{V} \\
    \dot{\mathcal{V}}	
    \end{bmatrix}
    \equiv \mathcal{L}(t)\,\mathcal{U}(t),
\end{equation}
where $\mathbb{O}$ and $\mathbb{I}$ are $2\times2$ zero and identity matrices, respectively, and $\mathcal{L}(t)$ is called the circuit Liouvillian matrix.
We also define $\mathcal{L}_0(t)$ for the uncoupled case (i.e., $\mathcal{H}_c=0$).
This form is more similar to the Schr\"odinger Eq. \eqref{schEQ} if one regards $i\mathcal{L}$ as the Hamiltonian.

For quantum LZT described by Eq. \eqref{schEQ}, the probability of state $|\psi(t)\rangle$ remaining in channel $i$ at time $t$ can be evaluated using the definition \cite{Zener_1932_PRSA}
\begin{eqnarray}\label{quP_channel}
	P_{i}(t) = \dfrac{|\psi_{i}(t)|^2}{|\psi_{1}(t)|^2+|\psi_{2}(t)|^2}~~~(i=1,2),
\end{eqnarray}
given the initial state $|\psi(-\infty)\rangle$ prepared only in channel $i$; correspondingly, the probability of tunneling to the other channel is $T_i(t)\equiv 1-P_i(t)$.

Following this definition, one may construct a similar quantity for circuit LZT using node voltages instead.
To this end, we should project the state vector $\mathcal{U}(t)$ onto four instantaneous eigenvectors $\mathcal{U}_{1,2}(t)$ and $\mathcal{U}^*_{1,2}(t)$ of the Liouvillian matrix $\mathcal{L}_0(t)$ of the uncoupled case, yielding
\begin{eqnarray}\label{eq:proj2channel}
    \mathcal{U}(t)&=&c_1(t)\,\mathcal{U}_1(t)+c_2(t)\,\mathcal{U}_2(t) +\text{c.c.}\notag\\
    &=&c_1(t)
    \begin{bmatrix}	
     1\\
     0\\
    i\omega_1\\
    0
    \end{bmatrix}
    +c_2(t)
    \begin{bmatrix}	
    0 \\
    1\\
    0\\
    i\omega_2	
    \end{bmatrix}
    +\text{c.c.},
\end{eqnarray}
where $c_{1,2}(t)$ are the complex coefficients that reflect the time evolution of node voltages through the relation $V_{1,2}(t)=2|c_{1,2}(t)|\cos \theta_{1,2}(t)$, with $\theta_{1,2}(t)$ being the arguments of $c_{1,2}(t)$.
Therefore, it seems that one can similarly define the remaining probability for channels as Eq. \eqref{quP_channel} by just replacing $\psi_{1,2}(t)$ with $c_{1,2}(t)$. 
However, there is one problem with this definition: Even for the uncoupled case, the modulus of $c_{2}(t)$ can increase with time [see the gray lines in Figs. \ref{fig2}(b)--\ref{fig2}(h)], which hints that the change in channel mode may stem from both the channel itself and the coupling to the other channel. 
Therefore, to faithfully evaluate the change only from the coupling, the remaining probability of channel $i$ at time $t$ for circuit LZT can be defined as
\begin{eqnarray}\label{Pt}
	P^{(c)}_{i}(t) = \dfrac{|c_{i}(t)|^2}{\big|c^{(r)}_{i}(t)\big|^2},
\end{eqnarray}
where the coefficient $c_i(t)$ is calculated by Eqs. \eqref{eq:1stode} and \eqref{eq:proj2channel} with initial values $\mathcal{U}(0)=[1,0,0,0]^T$ and $[0,1,0,0]^T$ for $i=1$ and $2$, respectively; the coefficient $c_i^{(r)}(t)$ for the uncoupled case is also calculated in the same way, using $\mathcal{L}_0(t)$ instead of $\mathcal{L}(t)$ in Eq. \eqref{eq:1stode}.
Apparently, taking $c_i^{(r)}(t)$ as the reference coefficient, the probability
$P^{(c)}_{i}(t)$ is bounded by values $0$ and $1$, which respectively correspond to no and all fractions of the channel mode remaining in node $i$ at time $t$; 
thus, the tunneling probability from channel $i$ to the other can be naturally defined as $T^{(c)}_{i}(t) \equiv 1-P^{(c)}_{i}(t)$. 
The definition in Eq. \eqref{Pt} can be regarded as a generalization of Eq. \eqref{quP_channel} to circuit systems.

The third row of Fig. \ref{fig2} shows the remaining probabilities of channels for different sweeping rates. 
No matter which node the initial voltages are prepared in, the remaining probabilities defined in Eq. \eqref{Pt} are the same and become larger as the sweeping rate increases. 
While a very slow or very fast sweep follows an adiabatic or uncoupled evolution, manifested by the fact that almost all or no fractions of the channel mode are transferred from the initial node to the other at late times, i.e., $P^{(c)}_i(t\rightarrow\infty)\sim 0$ or 1, as shown in Fig. \ref{fig2}(c) or Fig. \ref{fig2}(i), a moderate sweep can retain partial fractions of the node voltage in the initial nodes at late times, as shown in Fig. \ref{fig2}(f).  
These phenomena are consistent with the conventional LZT in quantum systems \cite{Landau_1932_PZS,Zener_1932_PRSA,Stuckelberg_1932_HPA,MajoranaZ_1932_IINC}.  

\section{Relation to quantum LZT} \label{sec:relation}

\begin{figure}[tb]
	\centering
	\includegraphics[width=1 \linewidth]{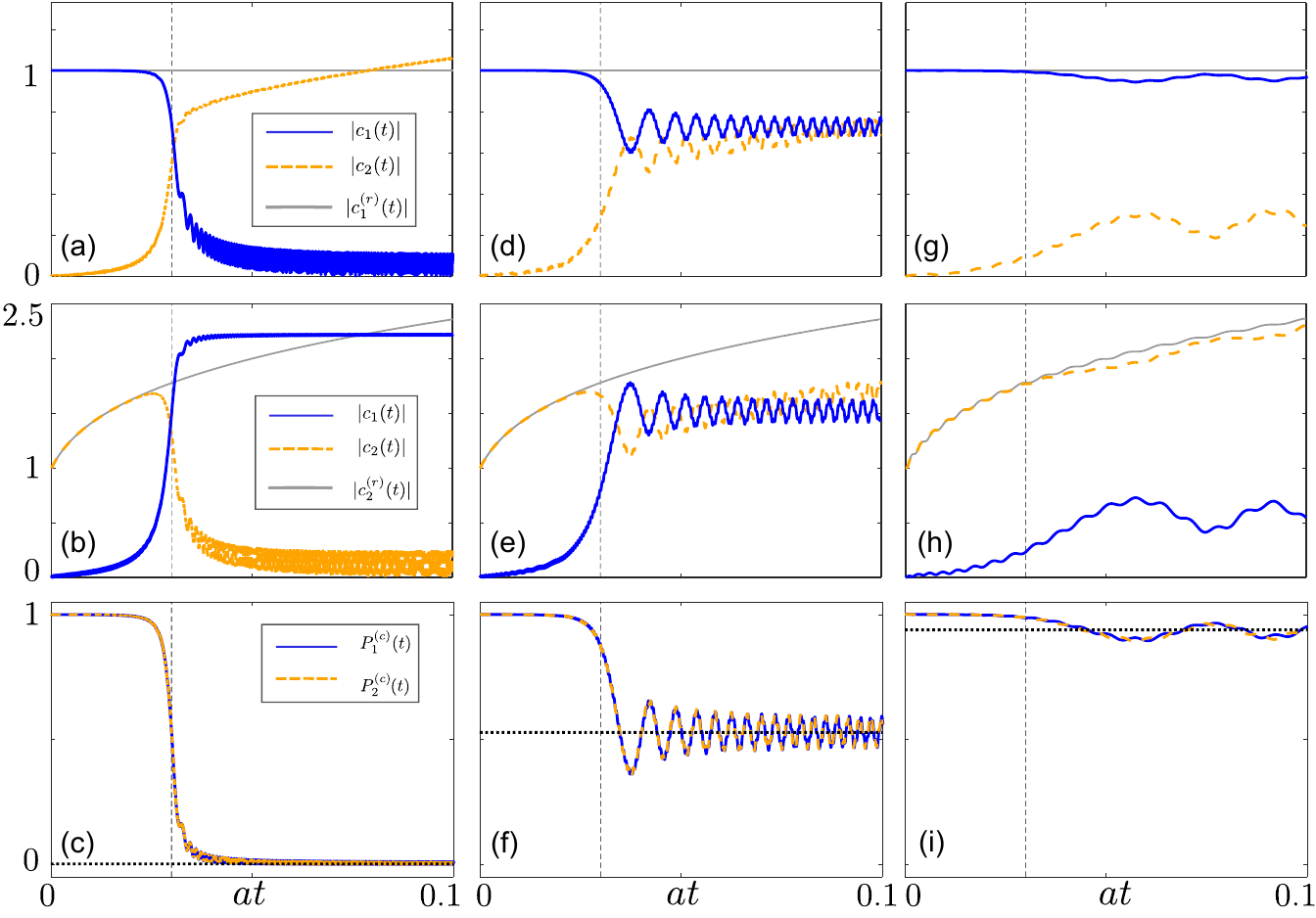}
	\caption{Nonreciprocal case with $v=1.25$. All other settings are the same as in Fig. \ref{fig2}. 
	}
	\label{fig3}
\end{figure}

In the quantum LZT described by Eq. \eqref{schEQ}, two important quantities are the instantaneous energy gap $\nu$ and the sweeping rate $\alpha$, which determine the probability of LZT through the characteristic dimensionless sweeping rate $\tilde\alpha\equiv\alpha/\nu^2$ in the analytical result $P_1(t\rightarrow\infty)=P_2(t\rightarrow\infty)=\exp{(-\pi/2\tilde\alpha)}$ \cite{Zener_1932_PRSA} (Appendix \ref{asec:LZT} reviews the derivation).

Although the differential equations of motion for circuit and quantum LZTs are different, the underlying physical nature may be the same.
To estimate the circuit LZT as its quantum counterpart, we linearize the frequency of channel 2 in time around the crossing time $t_c$ as
\begin{equation}\label{lin_w2}
    \omega_2\approx \sqrt{r}-\alpha'(t-t_c), 
\end{equation}
where
\begin{equation}\label{sweep_rate}
\alpha'=\frac{(r^2-1)^2}{2r^{5/2}}a\ge0
\end{equation}
denotes the sweeping rate of $\omega_2$ at $t=t_c$.
Thus, the circuit Hamiltonian \eqref{CcCl} can be approximated as
\begin{eqnarray}\label{lin_H}
    \mathcal{H}&\approx& 
    \begin{bmatrix}	
    ~r &~v \\
    ~v^{-1} & ~[\sqrt{r}-\alpha'( t-t_c)]^{2}	
    \end{bmatrix}\equiv \mathcal{H}_l.
\end{eqnarray}
Recalling the frequency gap $\Delta$ in Eq. \eqref{freq_gap}, we can similarly construct a dimensionless sweeping rate $\tilde\alpha'\equiv\alpha'/\Delta^2$, akin to its quantum counterpart, and ask whether the probabilities $P_i^{(c)}(t)$ at late times are somewhat close to the analytical value $\exp(-\pi/2\tilde\alpha')$ of the quantum LZT.
Figures \ref{fig2}(c), \ref{fig2}(f), and \ref{fig2}(i) show good agreement between the numerical results of circuit LZT and the analytical values from the perspective of quantum LZT, verifying our conjecture.
This means that the circuit LZT is not determined by the change rate of capacitors $a$, all of which are very small in Figs. \ref{fig2}(c), \ref{fig2}(f), and \ref{fig2}(i), but is determined by the sweeping rate $\alpha'$ and the frequency gap $\Delta$ near the crossing point $at_c$, i.e., by the dimensionless sweeping rate $\tilde\alpha'$, which characterizes the extent of nonadiabaticity of the circuit LZT.

To explain why circuit LZT satisfies the law of quantum LZT, we try to block-diagonalize $i\mathcal{L}$ as
\begin{equation}
    \mathcal{S}^{-1}(i\mathcal{L})\mathcal{S}=
    \begin{bmatrix}
    \mathbb{H}&\mathbb{O} \\	
    \mathbb{O}& -\mathbb{H}
    \end{bmatrix},
\end{equation}
where
\begin{equation}
\mathcal{S}=\frac{1}{\sqrt{2}}
    \begin{bmatrix}
    \mathbb{I}&\mathbb{I} \\	
    -i\mathbb{H}& i\mathbb{H}	
    \end{bmatrix},~~~~
\mathcal{S}^{-1}=\frac{1}{\sqrt{2}}
    \begin{bmatrix}
    \mathbb{I}& i\mathbb{H}^{-1}\\	
    \mathbb{I}& -i\mathbb{H}^{-1}	
    \end{bmatrix}
\end{equation}
are the time-dependent similarity matrix and its inverse matrix, respectively, and 
\begin{equation}\label{circuit_H}
\mathbb{H}=\frac{1}{\beta}
    \begin{bmatrix}
    \omega_1^2+\gamma& v \\	
    v^{-1} & \omega_2^2+\gamma	
    \end{bmatrix},~
\end{equation}
with $\gamma=\sqrt{\omega_1^2\omega_2^2-1}$ and $\beta=\sqrt{\omega_1^2+\omega_2^2+2\gamma}$, satisfying the relation $\mathbb{H}^2=\mathcal{H}$.
Under this similarity transformation, Eq. \eqref{eq:1stode} can be written in the form of a time-dependent Schr\"odinger equation:
\begin{equation}\label{unitary_transformation}
    i\,\dot{\mathcal{U}'}= \left\{
    \begin{bmatrix}	
    \mathbb{H} & \mathbb{O} \\
    \mathbb{O} & -\mathbb{H}	
    \end{bmatrix}
    -\frac{i}{2}
    \begin{bmatrix}	
    \mathbb{H}^{-1}\dot{\mathbb{H}} & -\mathbb{H}^{-1}\dot{\mathbb{H}} \\
    -\mathbb{H}^{-1}\dot{\mathbb{H}} & \mathbb{H}^{-1}\dot{\mathbb{H}}	
    \end{bmatrix}\right\}\mathcal{U}',
\end{equation}
where $\,\mathcal{U}'\equiv\mathcal{S}^{-1}\mathcal{U}$,
and the additional term with $\dot{\mathbb{H}}$ in the curly braces, stemming from the time dependence of $\mathcal{S}^{-1}$, makes the entire coefficient matrix not fully block diagonalized and introduces non-Hermiticity.

To further investigate Eq. \eqref{unitary_transformation}, we consider the linearized circuit Hamiltonian $\mathcal{H}_l$ in Eq. \eqref{lin_H} under the conditions
\begin{equation}
r\gg \max\{v,v^{-1}\}\sim O(1),~~~~ \sqrt{r}\gg \alpha'(t-t_c),
\end{equation}
where the first one is just the condition in Eq. \eqref{freq_gap} required by our circuit scheme, and the second one ensures that the linearized frequency $\omega_2$ in Eq. \eqref{lin_w2} is far above zero for the entire time evolution.
Then, to the first order of both $1/\sqrt{r}$ and $\alpha'(t-t_c)/\sqrt{r}$, Eq. \eqref{circuit_H} can be approximated as
\begin{equation}
\mathbb{H}\approx 
    \begin{bmatrix}
    \sqrt{r}& v/2\sqrt{r} \\	
    v^{-1}/2\sqrt{r} & \sqrt{r}-\alpha' (t-t_c)	
    \end{bmatrix},~
\end{equation}
and thus,
\begin{equation}\label{additional}
\mathbb{H}^{-1}\dot{\mathbb{H}}\approx -\frac{\alpha'}{\sqrt{r}}
    \begin{bmatrix}
    0& 0 \\	
    0 & 1	
    \end{bmatrix}. 
\end{equation}
Apparently, for the reciprocal case (i.e., $v=1$), the first term, which is Hermitian, in the curly braces of Eq. \eqref{unitary_transformation} represents two uncoupled quantum LZTs, described by $\pm\mathbb{H}$, both of which have the sweeping rate $\alpha'$ and the gap $1/\sqrt{r}$ (the approximate value of $\Delta$). 
According to the analytical results of quantum LZT \cite{Zener_1932_PRSA}, the probabilities of both at late times are just $\exp(-\pi/2\tilde{\alpha}')$ with
$\tilde{\alpha}'=\alpha'/\Delta^2\approx\alpha'r$, as we found in Fig. \ref{fig2}. 
The second term in the curly braces of Eq. \eqref{unitary_transformation} has a limited impact on the probability if $\alpha'/\sqrt{r}$ is small. However, this term introduces non-Hermiticity, primarily leading to the amplification of the channel-2 mode during the time evolution, which can be roughly expected from Eq. \eqref{additional}.

\section{Circuit LZT: nonreciprocal case}\label{sec_v}

From the analysis in Sec. \ref{sec:relation}, one can similarly understand the nonreciprocal case (i.e., $v\ne1$) of circuit LZT from two almost-decoupled nonreciprocal quantum LZTs, which are analytically solved in Ref. \cite{Torosov_Vitanov_2017_PRA} (a quick derivation can be found in Appendix \ref{Sec.0}).

Numerically, using Eq. \eqref{eq:1stode} with the same initial values of $\mathcal{U}(0)$ as in the reciprocal case, we can also plot the time evolutions of the coefficients $c_{1,2}(t)$ in Eq. \eqref{eq:proj2channel} and, thus, the remaining probabilities of channels $P^{(c)}_{1,2}(t)$ in Eq. \eqref{Pt}, as shown in Fig. \ref{fig3}. 
Compared with Fig. \ref{fig2}, although $|c_{1,2}(t)|$ are different, the probabilities $P^{(c)}_{1,2}(t)$ are identical to those of the reciprocal case, approaching the analytical value of $\exp(-\pi/2\tilde{\alpha}')$, where $\tilde{\alpha}'$ has the same values due to the identical instantaneous frequency spectrum for $v=1$ and $v\ne 1$, as shown in Fig. \ref{fig1}(b).

Through a time-independent similarity matrix
\begin{equation}
    \mathcal{R}=
    \begin{bmatrix}	
    \mathbb{R} & ~\mathbb{O} \\
    \mathbb{O} & ~\mathbb{R}	
    \end{bmatrix}~~
\text{with}~
    \mathbb{R}=
    \begin{bmatrix}	
    1 & ~0 \\
    0 & ~v^{-1}	
    \end{bmatrix},
\end{equation}
the differential equation \eqref{eq:1stode} for the nonreciprocal circuit LZT can be transformed into the reciprocal one as
\begin{equation}
    \frac{d}{dt}[\mathcal{R}^{-1}\mathcal{U}(t)]=
\mathcal{L}_r(t)[\mathcal{R}^{-1}\mathcal{U}(t)]
\end{equation}
where 
\begin{equation}
\mathcal{L}_r(t)\equiv \mathcal{R}^{-1}\mathcal{L}(t)\mathcal{R}
=\begin{bmatrix}
    \mathbb{O} & ~\mathbb{I} \\
    -\mathcal{H}_r & ~\mathbb{O}	
    \end{bmatrix},
\end{equation}
and
\begin{equation}
    \mathcal{H}_r \equiv \mathbb{R}^{-1}\mathcal{H}\mathbb{R}=
    \begin{bmatrix}	
    \omega_1^2 & ~1 \\
    1 & ~\omega_2^2	
    \end{bmatrix}
\end{equation}
are the circuit Liouvillian and Hamiltonian matrices for the reciprocal case, respectively.
Therefore, the solution to Eq. \eqref{eq:1stode} for the nonreciprocal case yields
\begin{eqnarray}
\mathcal{U}(t)&=&\mathcal{R}\left\{\hat{T}\exp\left[\int_0^t\mathcal{L}_r(\tau)d\tau\right]\right\}[\mathcal{R}^{-1}\mathcal{U}(0)],
\end{eqnarray}
where $\hat{T}$ is the time-ordering operator.
Apparently, the time evolution of the state for nonreciprocal and reciprocal cases differs only by a similarity transformation in the initial and final states.
From the expansion in Eq. \eqref{eq:proj2channel}, one can recognize that the similarity transformation does not affect the coefficient $c_i(t)$ when the initial state is prepared only in channel $i$ and suppresses or amplifies the other coefficient if $i=1$ or 2, as shown in Fig. \ref{fig3}. 
The state can be explicitly demonstrated as
\begin{eqnarray}\label{nr_c1}
    \mathcal{U}(t)&=&c_1(t)\,\mathcal{U}_1(t)+v^{-1}c_2(t)\,\mathcal{U}_2(t) +\text{c.c.}
\end{eqnarray}
for $\,\mathcal{U}(0)=[1,0,0,0]^T$, and 
\begin{eqnarray}\label{nr_c2}
    \mathcal{U}(t)&=&vc_1(t)\,\mathcal{U}_1(t)+c_2(t)\,\mathcal{U}_2(t) +\text{c.c.}
\end{eqnarray}
for $\,\mathcal{U}(0)=[0,1,0,0]^T$,
where $c_{1,2}(t)$ are the corresponding coefficients for the reciprocal case.
Thus, the remaining probability of channels $P_i^{(c)}(t)$ defined in Eq. \eqref{Pt} must be identical for nonreciprocal and reciprocal cases because the reference coefficients $c^{(r)}_{1,2}(t)$ also satisfy the laws of Eqs. \eqref{nr_c1} and \eqref{nr_c2}. 

It should be noted that, although the definitions of probability are different, our results are consistent with those in Ref. \cite{Torosov_Vitanov_2017_PRA} for nonreciprocal quantum LZT, 
but our definition, by removing the natural variations of the channel itself, ensures that both the remaining and tunneling probabilities do not exceed $1$, preserving the meaning of ``probability".

\section{Conclusions}\label{Sec:conclusion}
	
    In conclusion, we propose an electric circuit using time-varying capacitors to analyze the LZT phenomenon in classical systems. We find that, although the underlying differential equation of motion is quite different from the Schr\"odinger equation and the instantaneous frequency spectrum is not linear, based on our generalized definition for norm-unconserved systems, the probability still follows the laws of the LZT in quantum systems, which are codetermined by the linear sweeping rate and the gap.
    The deep relationship between the circuit LZT and its quantum counterpart can be established through a linearization and block-diagonalization process. 
    Our proposal provides a general method for simulating time-dependent quantum models using time-varying circuits, which has been lacking in previous studies.
    Given the increasing interest in the nonlinear non-Hermitian LZT \cite{Wang_Fu_2023_NJP} and the development of electric circuits for simulating nonlinear models \cite{Wang_Chong_2019_NC}, it is also possible to simulate more complicated LZT and other dynamical phenomena in nonlinear or interacting systems using circuits.
	
    \begin{acknowledgments}
        This work was supported by the National Key Research and Development Program of China (Grant No. 2022YFA1405304), the Guangdong Basic and Applied Basic Research Foundation (Grant No. 2024A1515010188), the Guangdong Provincial Quantum Science Strategic Initiative (Grant No. GDZX2204003), and the Startup Fund of South China Normal University.
    \end{acknowledgments}

    \section*{DATA AVAILABILITY}
        The data that support the findings of this article are openly available \cite{data}.

    \appendix
    	
\section{Derivation of the differential equation of the circuit}\label{Anote2}

    Here, we present the derivation for the differential Eqs. \eqref{EQvoltage} and \eqref{ddv} of the circuit shown in Fig. \ref{fig1}(a). 
    
    Considering the characteristic equations
    \begin{eqnarray}
		V=IR,~~~~V=L\dot{I},~~~~\dot{C}V+C\dot{V}=I,
	\end{eqnarray}
for resistors, inductors, and time-varying capacitors, respectively, where $V$ is the voltage drop across the element following the direction of current $I$, 
we apply Kirchhoff's current law to nodes $1$ and $2$, yielding
	\begin{eqnarray}\notag
		0&=&-(C_1+C_3)\ddot{V}_1+C_3\ddot{V}_2-(2\dot{C}_1+2\dot{C}_3+R_1^{-1}+R_3^{-1}) \\ \notag
        &&\dot{V}_1+(2\dot{C}_3+R_3^{-1})\dot{V}_2 -(\ddot{C}_1+\ddot{C}_3+L_1^{-1})V_1+\ddot{C}_3 V_2, \\\notag
        0&=&-(C_2+C_3)\ddot{V}_2+C_3\ddot{V}_1-(2\dot{C}_2+2\dot{C}_3+R_2^{-1}+R_3^{-1}) \\ \notag
        &&\dot{V}_2+(2\dot{C}_3+R_3^{-1})\dot{V}_1 -(\ddot{C}_2+\ddot{C}_3+L_2^{-1})V_2+\ddot{C}_3 V_1. \\
	\end{eqnarray}
One can write them in a compact form:
	\begin{equation}\label{circuit_eq}
		\mathcal{C}_C(t)\ddot{\mathcal{V}}(t)+\mathcal{C}_R(t)\dot{\mathcal{V}}(t)+\mathcal{C}_L(t)\mathcal{V}(t)=0, 
	\end{equation}
where the vector $\mathcal{V}(t) = [V_1,V_2]^{\rm T}$ and the coefficient matrices
\begin{equation}\label{coefficientMatrixAP}
\begin{aligned}
	\mathcal{C}_C(t) &= 
        \begin{bmatrix}
        -C_3-C_1&C_3 \\
        C_3&-C_3-C_2
        \end{bmatrix},\\
	\mathcal{C}_R(t) &=
        \begin{bmatrix}
        -C_R^{(1)}& 2\dot{C}_3+R_3^{-1}\\
        2\dot{C}_3+R_3^{-1}&-C_R^{(2)}
        \end{bmatrix},\\
	\mathcal{C}_L(t) &=
        \begin{bmatrix}
        -C_L^{(1)} & \ddot{C}_3\\
        \ddot{C}_3 &-C_L^{(2)}
        \end{bmatrix},
\end{aligned}
\end{equation}
with $C_R^{(i)} = 2(\dot{C}_i+\dot{C}_3)+R_i^{-1}+R_3^{-1}$ and $C_L^{(i)} = \ddot{C}_i+\ddot{C}_3+L_i^{-1}$.

Following the parametrization in Eq. \eqref{param}, we can obtain
	\begin{eqnarray}\notag
		\mathcal{C}_C(t) &=& C_0
		\begin{bmatrix} 
        -(at+b+r)/vr^2 & ~(at + b)/r \\ 
        (at + b)/r&	~-v(at + b) 
		\end{bmatrix}, \\
		\mathcal{C}_L(t) &=& -\frac{1}{L_0}
		\begin{bmatrix} 1/v&~0 
        \\ 0&	~v 
		\end{bmatrix}, ~~~~\mathcal{C}_R(t) = 0,
	\end{eqnarray}
    and the differential equation \eqref{circuit_eq} can be reduced to
    	\begin{equation}
		\ddot{\mathcal{V}}(t)=-\mathcal{H}(t)\mathcal{V}(t), 
	\end{equation}
where 
	\begin{eqnarray}
		\mathcal{H}(t) &=& \mathcal{C}_C^{-1}(t)\mathcal{C}_L(t)\notag\\
        &=&\frac{1}{L_0C_0}
		\begin{bmatrix}	
        ~r~&~v \\~~	v^{-1}&~~~	{r}^{-1}+({at+b})^{-1}	
        \end{bmatrix},
	\end{eqnarray}
    obtained using the inverse of $\mathcal{C}_C(t)$:
    	\begin{eqnarray}
		\mathcal{C}_C^{-1}(t) = -\frac{1}{C_0}
        \begin{bmatrix}
        ~	vr  &~1~\\~1~& ~~~{(vr)}^{-1}+{[v(at+b)]}^{-1} 
        \end{bmatrix}.
	\end{eqnarray}

	To ensure the stability of the circuit, the eigenfrequencies $\omega_\pm$ must be real, which requires that $\omega_\pm^2$, as the eigenvalues of $\mathcal{H}(t)$, must both be positive. To this end, we just need $at + b > 0$.
    The proof is as follows:
    Two eigenvalues of $\mathcal{H}(t)$ read
    \begin{eqnarray}
		\lambda_\pm=\frac{r+r^{-1}+T^{-1}\pm\sqrt{(r-r^{-1}-T^{-1})^2+4}}{2L_0C_0},
	\end{eqnarray}
    where $T=at+b$.
    Because all parameters are real, to have all eigenvalues nonnegative, we only need to require the smaller eigenvalue to be nonnegative, i.e.,
    \begin{eqnarray}
		r+r^{-1}+T^{-1}\ge\sqrt{(r-r^{-1}-T^{-1})^2+4},
	\end{eqnarray}
    which can be further reduced by squaring both sides, yielding
    \begin{eqnarray}
		r(r^{-1}+T^{-1})\ge1.
	\end{eqnarray}    
Finally, we have the condition $T=at+b>0$ since $r>0$.
Therefore, we consider the dynamical process from $t=0$ to $+\infty$, with $b$ being a small positive number to prevent divergence at $t=0$.

    This circuit design for generating $\mathcal{H}(t)$ relies on three time-varying capacitors $C_{1,2,3}$, which vary linearly with time (see Appendix \ref{AnoteTVC} for details). 
This produces the constant parts $r^{\pm 1}$ of the diagonal terms, along with the time-dependent part $(at+b)^{-1}$. 
Nonreciprocal coupling $v^{\pm 1}$ in the off-diagonal terms is implemented by introducing two inductors $L_{1,2}$.
Using linearly time-varying capacitors may be a better scheme than using time-varying inductors or resistors, which could involve an integro-differential equation instead.

\section{Time-varying capacitors}\label{AnoteTVC}

    \begin{figure}[tb]
		\includegraphics[width=1\linewidth]{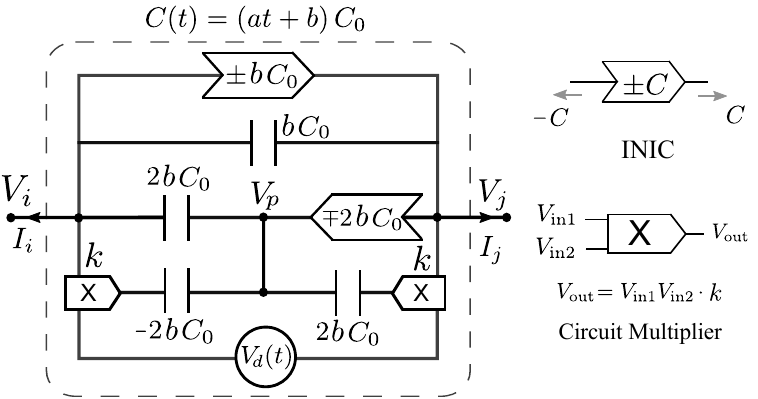}
		\caption{The schematic diagram for the linearly time-varying capacitance $C(t)=(a t + b)C_0$ (the left panel circled out by the dashed rectangle). The right panel demonstrates the INIC and the circuit multiplier. The detailed description can be found in the text.}
        \label{fig4}
    \end{figure}

	One may realize the time-varying capacitors directly by changing the plates of capacitors. 
    Here, we propose a delicate scheme for the linearly time-varying capacitance $C(t)=(a t + b)C_0$ between nodes $i$ and $j$, as shown in Fig. \ref{fig4}. 
    We use the impedance converters through current inversion (INIC) \cite{Hofmann_Thomale_2019_PRL,Helbig_Thomale_2020_NP} to realize the capacitance, the sign of which depends on the direction. The negative capacitance $-2bC_0$ in the circuit can be realized according to Ref.  \cite{Schindler_Kottos_2012_JoPA}.
    We also use the circuit multiplier \cite{Stegmaier_Upreti_2024_PRR,Stegmaier_Thomale_2024_arxiv,Zhang_Zhang_2025_NC}, which multiplies the node voltage $V_{i,j}$ and the external voltage source $V_d(t) = at V_0$ ($V_0$ is a reference voltage) by an amplification factor $k=1/b V_0$.

	Here, we consider the ideal INICs and multipliers, and the currents flowing into nodes $i$ and $j$ can be expressed as 
	\begin{eqnarray}
	I_i &=& 2b C_0 (\dot{V_p}-\dot{V_i}) ,\label{Ii}\\
        I_j &=& -2b C_0 (\dot{V_p}-\dot{V_j}) + 2b C_0 (\dot{V_i}-\dot{V_j}).\label{Ij}
	\end{eqnarray} 
	According to the zero net current flowing into the node $p$, we obtain
	\begin{eqnarray}
	0&=& 2bC_0(\dot{V_i}-\dot{V_p}) + 2bC_0(\dot{V_j}-\dot{V_p}) \\
        && - 2bC_0 k\frac{d}{dt}(V_i V_d-V_p)+  2bC_0 k\frac{d}{dt}(V_j V_d-V_p),\notag
	\end{eqnarray}
    where we use the sign-inverted property of the INIC and the multiplication property of the circuit multiplier, as shown in Fig. \ref{fig4}. 
    Then we have
	\begin{equation}\label{dvp}
             \dot{V_p} = \frac{1}{2}\big[(kV_d+1)\dot{V_j} - (kV_d-1)\dot{V_i} + k\dot{V_d}(V_j-V_i)\big].
	\end{equation}
	Substituting Eq. \eqref{dvp} into Eqs. \eqref{Ii} and \eqref{Ij}, and setting the amplification factor $k=1/b V_0$ and the external voltage source $V_d(t) = at V_0$, we obtain
	\begin{eqnarray}
	I_i = (a t +b)C_0(\dot{V_j}-\dot{V_i})+ a C_0\left({V_j}-{V_i}\right)=-I_j.
	\end{eqnarray}
    Compared with the characteristic equation of time-varying capacitors $I=C\dot{V}+\dot{C}V$, one can recognize that the effective capacitance $C(t)=(at+b)C_0$ has been achieved.
     
\section{Circuit LZT with different initial values} \label{asec:nonzero_dotv}

\begin{figure}[tb]
	\centering
	\includegraphics[width=1 \linewidth]{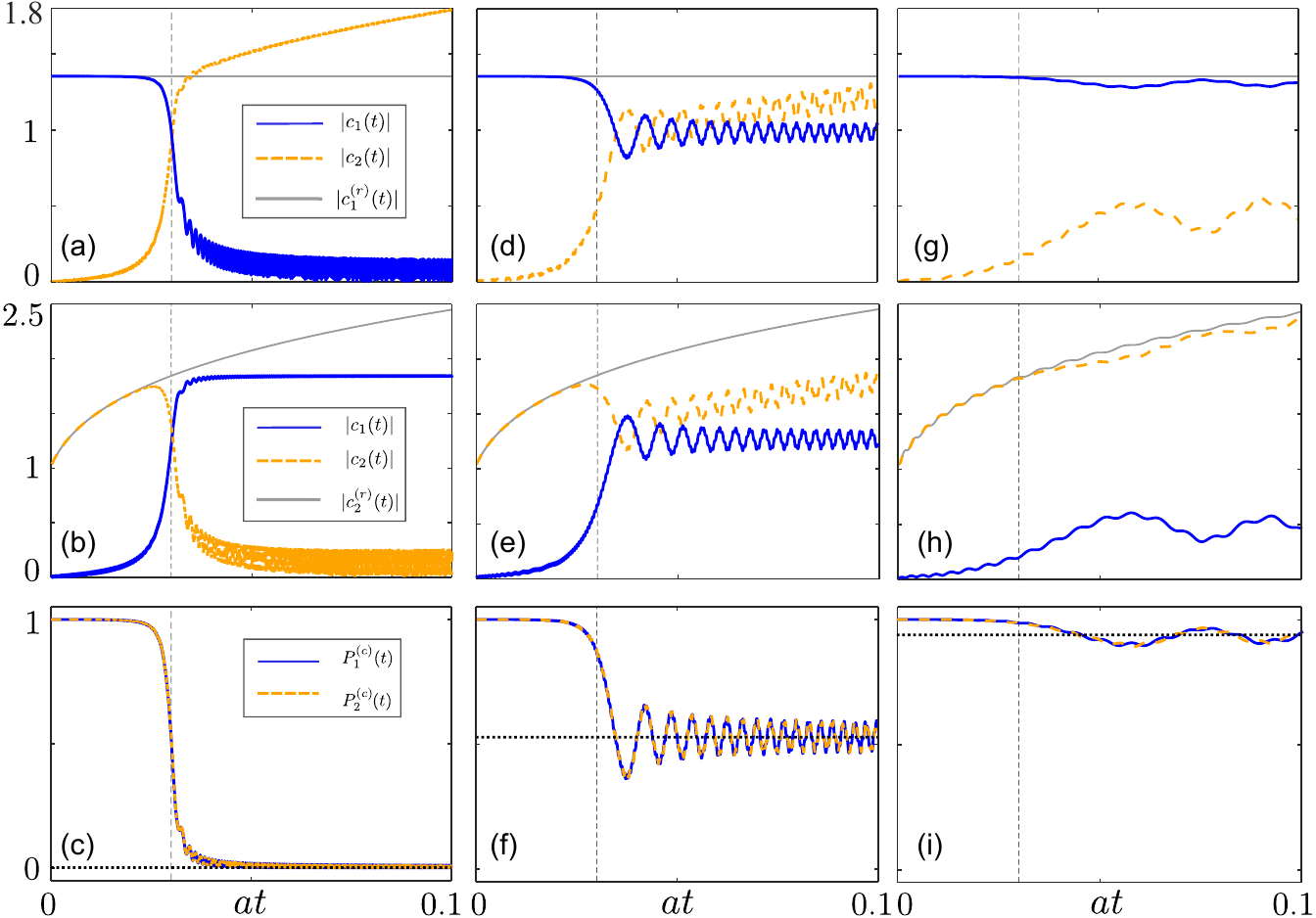}
	\caption{Reciprocal ciruit LZT with initial values $\mathcal{U}^T(0)=[1,0,5,0]$ and 
    $[0,1,0,5]$ instead of $[1,0,0,0]$ and 
    $[0,1,0,0]$ in Fig. \ref{fig2}, respectively. All other settings are the same as in Fig. \ref{fig2}.
	}
	\label{fig5}
\end{figure}

In Fig. \ref{fig5}, we demonstrate the reciprocal circuit LZT with initial values $\mathcal{U}^T(0)=[1,0,5,0]$ and 
    $[0,1,0,5]$ instead of $[1,0,0,0]$ and 
    $[0,1,0,0]$ in Fig. \ref{fig2}, respectively, while keeping the other settings the same.  
    We can see that, except for the overall amplification of $|c_{1,2}(t)|$ and $|c_{1,2}^{(r)}(t)|$, the final remaining probabilities are identical to those in Fig. \ref{fig2}.
    The reason can be explained as follows.

    According to Eq. \eqref{eq:proj2channel}, the initial values can be expressed as
	\begin{eqnarray}
		V_{1,2}(0)&=&2|c_{1,2}(0)|\cos\theta_{1,2}(0),\notag\\
        \dot{V}_{1,2}(0)&=&-2\omega_{1,2}(0)|c_{1,2}(0)|\sin\theta_{1,2}(0),
    \end{eqnarray}
where the coefficients $c_{1,2}(t)=|c_{1,2}(t)|\exp[i\theta_{1,2}(t)]$ with the subscripts 1,2 labeling the initial channels.
Then we can obtain
	\begin{eqnarray}\label{multiplier}
		|c_{1,2}(0)|=\frac{1}{2}\sqrt{V_{1,2}^2(0)+\frac{\dot{V}_{1,2}^2(0)}{\omega_{1,2}^2(0)}}.
    \end{eqnarray}
Apparently, different initial values of $V_{1,2}(0)$ and $\dot{V}_{1,2}(0)$ correspond to different $|c_{1,2}(0)|$ and $\theta_{1,2}(0)$, which generate only a common multiplier of Eq. \eqref{multiplier} for the successive dynamics of $|c_{1,2}(t)|$ and $|c_{1,2}^{(r)}(t)|$, following Eq. \eqref{eq:proj2channel}. 
Thus, the remaining probability defined in Eq. \eqref{Pt} is irrelevant to the initial values.
Likewise, one can expect the same conclusion for the nonreciprocal case (i.e., $v\ne 1$).

For experiments, the initial voltage $V(0)$ can be implemented through techniques such as pulsed power application for node accessibility \cite{Lang_Liang_2021_PRB}, while the nonzero voltage change $\dot{V}(0)$ can be achieved simply by evolving the circuit over a certain period while keeping the time-varying capacitors $C_{1,2,3}$ unchanged at their initial values.

\section{Recap of analytical results of the reciprocal quantum LZT}  \label{asec:LZT}

    In this section, we review the analytical derivation of the reciprocal quantum LZT \cite{Zener_1932_PRSA}. 
    The time-dependent Schr\"odinger equation of a general form can be expressed as ($\hbar=1$)
	\begin{equation}\label{id/dt*f=H*f}
		i\frac{d}{dt}
        \begin{bmatrix}
			\psi_1(t)\\
			\psi_2(t)\\            
        \end{bmatrix}
       =H_r(t)
        \begin{bmatrix}
			\psi_1(t)\\
			\psi_2(t)\\            
        \end{bmatrix},
	\end{equation}
    where 
    \begin{equation}\label{gel_H}
        H_r(t)=\frac{1}{2}
       \begin{bmatrix}
			\epsilon+\alpha_1 t& \nu\\
			\nu& -\epsilon-\alpha_2 t\\
		\end{bmatrix}
    \end{equation}
    is the time-dependent Hamiltonian matrix with the reciprocal inter-channel coupling $\nu$,
    the changing rates $\alpha_{1,2}/2$, and the reference energies $\pm\epsilon/2$ of the two channel potentials; the amplitudes $\psi_{1,2}(t)$ are normalized as $|\psi_1(t)|^2+|\psi_2(t)|^2=1$.
    Without loss of generality, the parameters $\alpha_{1,2}$, $\epsilon$, and $\nu$ are assumed to be positive in the following.  
    
    Considering the solution to Eq. \eqref{id/dt*f=H*f} without the coupling (i.e., $\nu=0$), we can set the amplitudes in the following forms:
	\begin{eqnarray}
			\psi_1(t) &\equiv& a(t)\exp\left[-i(\frac{\alpha_1 t^2}{4} +\frac{\epsilon t}{2})\right],\notag\\
			\psi_2(t) &\equiv& b(t)\exp\left[i(\frac{\alpha_2 t^2}{4} +\frac{\epsilon t}{2})\right].	 
            \label{eq:Atoa}
	\end{eqnarray}
    Substituting them into Eq. (\ref{id/dt*f=H*f}), we obtain the first-order differential equations for $a(t)$ and $b(t)$
	\begin{equation}
	    \left\{
        \begin{aligned}
        	&~i\dot a(t)=\frac{\nu}{2} \exp\left[ i(\frac{\alpha t^2}{2}+\epsilon t)\right] b(t)\\
			&~i\dot b(t)=\frac{\nu}{2} \exp\left[-i(\frac{\alpha t^2}{2}+\epsilon t)\right] a(t)
        \end{aligned}
        \right.~~,
        \label{a and b}
        \end{equation}
    where $\alpha=(\alpha_1+\alpha_2)/2$. 
    By eliminating $a(t)$, a second-order differential equation for $b(t)$ is obtained
	\begin{eqnarray}
		\ddot b(t)+i(\alpha t+\epsilon)\dot b(t)+\frac{\nu^2}{4} b(t) = 0
		\label{oneeq_1}.
	\end{eqnarray}
    To solve this equation, we make the following variable and function substitutions in Eq. \eqref{oneeq_1}
	\begin{equation}
		z=\sqrt{\alpha}\exp\left(-i\frac{\pi}{4}\right)\left( t+\frac{\epsilon}{\alpha}\right),~~~~Y(z)=\exp\left(-\frac{z^2}{4}\right)b(t),
        \label{subsiti}
	\end{equation}
    yielding the standard form of the Weber equation \cite{Whittaker_Watso_2021_CUP}
	\begin{eqnarray}
		\frac{d^2}{dz^2}Y(z)+\left( \mu+\frac{1}{2} -\frac{z^2}{4}\right) Y(z) = 0,
        \label{eq:Weber}
	\end{eqnarray}
    where $\mu= i\nu^2/4\alpha$. 
    The parabolic cylinder functions (or Weber functions) $D_\mu(\pm z)$ and $D_{-\mu - 1}(\pm iz)$ are linearly independent solutions to this equation.
    To analyze the behaviors of LZT in the limits of $t\rightarrow\pm \infty$, we resort to the asymptotic expansions of the Weber function for $|z|\rightarrow \infty$ in the complex plane \cite{Whittaker_Watso_2021_CUP}
    \begin{eqnarray}
        D_{\mu}(z)&\sim& \exp\left(-\frac{z^2}{4}\right)z^{\mu}\sum_{s=0}^{\infty}\frac{(-1)^s\Gamma(-\mu+2s)}{(2z^2)^ss!\Gamma(-\mu)},\notag\\
        &&\left(|\arg z|<\frac{3\pi}{4}\right),
        \label{eq:asymp1}
    \end{eqnarray}
    and
    \begin{eqnarray}
	&&D_\mu(z)\sim \exp\left(-\frac{z^2}{4}\right)z^{\mu}\sum_{s=0}^{\infty}\frac{(-1)^s\Gamma(-\mu+2s)}{(2z^2)^ss!\Gamma(-\mu)}
    \nonumber \\
    &&~~~~+\frac{\sqrt{2\pi}}{\Gamma(-\mu)}\exp\left(\frac{z^2}{4}\right)(-z)^{-\mu-1}\sum_{s=0}^{\infty}\frac{\Gamma(\mu+1+2s)}{(2z^2)^ss!\Gamma(\mu+1)},
    \nonumber\\
    &&~~~~\left(\frac{\pi}{4}<\pm \arg z<\frac{5\pi}{4}\right),
    \label{eq:asymp2}
    \end{eqnarray}
    where $\Gamma(\mu)$ is the Gamma function.
    
    Here, we set the initial condition that only the low-energy level is occupied at $t\rightarrow -\infty$, yielding the corresponding boundary conditions of Eq. \eqref{eq:Weber}
    \begin{eqnarray}	    
			0&=&\lim_{t\rightarrow -\infty}\psi_2(t)=\lim_{z\rightarrow \infty e^{i\frac{3\pi}{4}}}Y(z),
            \label{eq:initial_cond1}\\
        	1&=&\lim_{t\rightarrow -\infty}|\psi_1(t)|=\lim_{z\rightarrow \infty e^{i\frac{3\pi}{4}}}=\frac{2\sqrt{\alpha }}{\nu}\left|\frac{dY(z)}{dz}+\frac{z}{2}Y(z)\right|.
            \nonumber\\
		\label{eq:initial_cond2}
    \end{eqnarray}
    For the condition in Eq. \eqref{eq:initial_cond1}, according to the asymptotic expansions in Eqs. \eqref{eq:asymp1} and \eqref{eq:asymp2}, we find that only the solution $D_{-\mu-1}(-iz)$ tends to zero for $|z|\rightarrow \infty$, i.e., 
    \begin{eqnarray}
	&&\lim_{z\rightarrow\infty \exp\left(i\frac{3\pi}{4}\right)}D_{-\mu-1}(-iz) \notag\\
    &\approx& \lim_{z\rightarrow\infty \exp\left(i\frac{3\pi}{4}\right)}\exp\left(\frac{z^2}{4}\right)(-iz)^{-\mu-1}
    \nonumber \\
    &=&\lim_{|z|\rightarrow\infty} \exp\left(\frac{\pi \nu^2}{16\alpha}\right)\exp\left(-i\frac{|z|^2+\pi}{4}\right)|z|^{-i\frac{\nu^2}{4\alpha}-1}
  \approx 0.\notag\\
    \end{eqnarray}
    Thus, the solution to Eq. \eqref{eq:Weber} satisfying the condition in Eq. \eqref{eq:initial_cond1} can be written as
    \begin{eqnarray}
		Y(z)=MD_{-\mu - 1}(-iz),
		\label{Y_z 2}
    \end{eqnarray}
    where $M$ is the coefficient determined by the condition in Eq. \eqref{eq:initial_cond2}
    \begin{eqnarray}	    
        	1&=&\lim_{z\rightarrow \infty e^{i\frac{3\pi}{4}}}\frac{2\sqrt{\alpha}\left|M\right|}{\nu}\left|\frac{dD_{-\mu - 1}(-iz)}{dz}+\frac{z}{2}D_{-\mu - 1}(-iz)\right|
            \nonumber\\
            &=&\lim_{z\rightarrow \infty e^{i\frac{3\pi}{4}}}\frac{2\sqrt{\alpha}\left|M\right|}{\nu}
            \nonumber\\
            &&\times\left|i(\mu+1)D_{-\mu - 2}(-iz)+zD_{-\mu - 1}(-iz)\right|
            \nonumber\\
            &\approx&\lim_{z\rightarrow \infty \exp\left(i\frac{3\pi}{4}\right)}\frac{2\sqrt{\alpha}\left|M\right|}{\nu}\left|\exp\left(\frac{z^2}{4}\right)(-iz)^{-\mu-2}\right|
            \left|\mu+1-z^2\right|
            \nonumber\\
            &=& \lim_{|z|\rightarrow \infty }\frac{2\sqrt{\alpha}\left|M\right|}{\nu |z|^{2}}
            \left|\frac{\nu^2}{4\alpha}+|z|^2-i\right| e^{\frac{\pi \nu^2}{16\alpha}}
            \nonumber\\
            &\approx& \frac{2\sqrt{\alpha}|M|}{\nu}e^{\frac{\pi \nu^2}{16\alpha}},
    \end{eqnarray}
    i.e., 
    \begin{eqnarray}
		|M|=\frac{\nu}{2\sqrt{\alpha}}\exp\left(-\frac{\pi \nu^2}{16\alpha}\right).
    \end{eqnarray}
\begin{widetext}
In the above derivation, we use the recurrence relation for the Weber function:     
	\begin{eqnarray}
		\frac{d}{dz}D_\mu(z)+\frac{z}{2}D_\mu(z)-\mu D_{\mu - 1}(z)=0.
	\end{eqnarray}

    Now we consider the late-time behavior of LZT in the limit of $t\rightarrow\infty$, i.e., the $Y(z)$ for infinite $z$ along the direction $\exp(-i\pi/4)$ in the complex plane.
    From the asymptotic expansion in Eq. \eqref{eq:asymp2}, we have
    \begin{eqnarray}
        \lim_{z\rightarrow\infty \exp\left(-i\frac{\pi}{4}\right)}|Y(z)|^2
        &=&\lim_{z\rightarrow\infty \exp\left(-i\frac{\pi}{4}\right)}|M|^2|D_{-\mu - 1}(-iz)|^2
        \nonumber\\
        &\approx& |M|^2\lim_{z\rightarrow\infty \exp\left(-i\frac{\pi}{4}\right)}\left|\exp\left(\frac{z^2}{4}\right)(-iz)^{-\mu-1}
        +\frac{\sqrt{2\pi}}{\Gamma(\mu+1)}\exp\left(-\frac{z^2}{4}\right)(iz)^{\mu}\right|^2 
        \nonumber\\
        &=&\frac{\nu^2}{4\alpha}\exp\left(-\frac{\pi \nu^2}{8\alpha}\right)\lim_{|z|\rightarrow\infty}\Big|\exp\left(-\frac{3\pi \nu^2}{16\alpha}\right)\exp\left(i\frac{3\pi-|z|^2}{4}\right)|z|^{-i\frac{\nu^2}{4\alpha}-1}
        \nonumber\\
        &&+\frac{\sqrt{2\pi}}{\Gamma(i\frac{\nu^2}{4\alpha}+1)}\exp\left(-\frac{\pi \nu^2}{16\alpha}\right)\exp\left(i\frac{|z|^2}{4}\right)|z|^{i\frac{\nu^2}{4\alpha}}\Big|^2
        \nonumber\\
        &\approx&\frac{\pi \nu^2\exp\left(-\frac{\pi \nu^2}{4\alpha}\right)}{2\alpha \Gamma(1+i\frac{\nu^2}{4\alpha})\Gamma(1-i\frac{\nu^2}{4\alpha})}
        \nonumber\\
        &=&1-\exp\left(-\frac{\pi \nu^2}{2\alpha}\right),
    \end{eqnarray}
    where we use the relation of the Gamma function:
	\begin{eqnarray}
		\Gamma(1 + \mu)\Gamma(1 - \mu)=\frac{\pi \mu}{\sin\pi \mu}.
	\end{eqnarray}
    
	Finally, according to Eqs. \eqref{eq:Atoa} and \eqref{oneeq_1}, we have
\begin{eqnarray}
		\lim_{t\rightarrow \infty}|\psi_2(t)|^2&=&\lim_{z\rightarrow\infty \exp\left(-i\frac{\pi}{4}\right)}\left|\exp\left(\frac{z^2}{4}\right)Y(z)\right|^2
        \approx 1-\exp\left(-\frac{\pi \nu^2}{2\alpha}\right),
        \nonumber\\
		\lim_{t\rightarrow \infty}| \psi_1(t) |^2&=&1-\lim_{t\rightarrow \infty}| \psi_2(t) |^2\approx \exp\left(-\frac{\pi \nu^2}{2\alpha}\right),
	\end{eqnarray}
    that is, the LZT from the low level to the high level (i.e., the remaining probability in channel 1) is
	\begin{eqnarray}
		P_{1}(t\rightarrow +\infty)=\lim_{t\rightarrow \infty}| \psi_1(t) |^2\approx \exp\left(-\frac{\pi \nu^2}{2\alpha}\right)
	\end{eqnarray}    
    
    Likewise, if we take a different initial condition in which only the high energy level is occupied at $t\rightarrow -\infty$, similar conclusions can be obtained using a similar derivation,
    i.e., 
        \begin{equation}
		\lim_{t\rightarrow \infty}|\psi_1(t)|^2\approx 1-\exp\left(-\frac{\pi \nu^2}{2\alpha}\right),~~~~
		\lim_{t\rightarrow \infty}| \psi_2(t) |^2\approx \exp\left(-\frac{\pi \nu^2}{2\alpha}\right),   
        \end{equation}
    and thus, 
    the LZT from the high level to the low level (i.e., the remaining probability in channel 2) is
	\begin{equation}
		P_{2}(t\rightarrow +\infty)=\lim_{t\rightarrow \infty}| \psi_2(t) |^2\approx \exp\left(-\frac{\pi \nu^2}{2\alpha}\right)=P_{1}(t\rightarrow +\infty).\label{ThlTlh}
	\end{equation}    
	It is worth noting that the results only concern the modulus of the solution at late times, leaving the phase undetermined, which depends on the phase of the initial state and the time at which the final state is measured.
\end{widetext}

\section{A quick derivation for nonreciprocal quantum LZT}\label{Sec.0}

For the nonreciprocal quantum LZT, which is described by
    \begin{equation}\label{nr_LZT}
		i\frac{d}{dt}
        \begin{bmatrix}
			\psi_1(t)\\
			\psi_2(t)           
        \end{bmatrix}
       =\frac{1}{2}
       \begin{bmatrix}
			\epsilon+\alpha_1 t& \nu e^\theta\\
			\nu e^{-\theta}& -\epsilon-\alpha_2 t
		\end{bmatrix}
        \begin{bmatrix}
			\psi_1(t)\\
			\psi_2(t)           
        \end{bmatrix},
	\end{equation}
    where the parameters are the same as those in Eq. \eqref{id/dt*f=H*f}, except that the interchannel couplings are nonreciprocal, as characterized by a real parameter $\theta$. Notably, the norm $|\psi_1(t)|^2+|\psi_2(t)|^2$ is not conserved, unlike the reciprocal case (i.e., $\theta=0$).

    A quick result can be obtained if one realizes the relationship between the nonreciprocal and reciprocal quantum LZTs through a similarity transformation
    \begin{equation}
		S=
        \begin{bmatrix}
			1 & 0\\
			0 & e^{-\theta}\\            
        \end{bmatrix},
	\end{equation}
    yielding
    \begin{equation}
		i\frac{d}{dt}\left(S^{-1}
        \begin{bmatrix}
			\psi_1(t)\\
			\psi_2(t)\\            
        \end{bmatrix}\right)
       =H_r(t)\left(S^{-1}
        \begin{bmatrix}
			\psi_1(t)\\
			\psi_2(t)\\            
        \end{bmatrix}\right),
	\end{equation}
    where 
\begin{equation}
H_r(t)= S^{-1}\left(\frac{1}{2}
       \begin{bmatrix}
			\epsilon+\alpha_1 t& \nu e^\theta\\
			\nu e^{-\theta}& -\epsilon-\alpha_2 t\\
		\end{bmatrix}\right)S,
\end{equation}
is just the reciprocal Hamiltonian matrix in Eq. \eqref{gel_H}.
Therefore, the solution to Eq. \eqref{nr_LZT} for the nonreciprocal case yields
\begin{equation}
        \begin{bmatrix}
			\psi_1(t)\\
			\psi_2(t)\\            
        \end{bmatrix}
        =S\hat{T}\exp\left[-i\int_{t_0}^{t} H_r(\tau)d\tau\right]
        \left(S^{-1}
        \begin{bmatrix}
			\psi_1(t_0)\\
			\psi_2(t_0)\\            
        \end{bmatrix}
        \right),
\end{equation}
where $\hat{T}$ is the time-ordering operator.
Apparently, the time evolution of the state for nonreciprocal and reciprocal cases differs only by a similarity transformation in the initial and final states.
Therefore, with the aid of the results for the reciprocal case, one can immediately obtain the probability of LZT from the low level to the high level (i.e., the remaining probability in channel 1)
	\begin{eqnarray}
		P_{1}(t\rightarrow +\infty)=\lim_{t\rightarrow \infty}| \psi_1(t) |^2\approx \exp\left(-\frac{\pi \nu^2}{2\alpha}\right)
	\end{eqnarray}    
and the probability from the high level to the low level (i.e., the remaining probability in channel 2) is
	\begin{equation}
		P_{2}(t\rightarrow +\infty)=\lim_{t\rightarrow \infty}| \psi_2(t) |^2\approx \exp\left(-\frac{\pi \nu^2}{2\alpha}\right)=P_{1}(t\rightarrow +\infty).
	\end{equation}    
The results are identical to those of the reciprocal quantum LZT.

Note that the results using the definition in Eq. \eqref{Pt} with the reference state are the same as the results demonstrated here because the reference state is norm-conserving when the coupling is turned off.

\bibliography{ref}
\end{document}